\documentclass[12pt]{article}
\usepackage{epsfig,graphicx}
\title{Single eta  production in heavy quarkonia: breakdown of multipole expansion
}
\author{ Yu.A.Simonov, A.I.Veselov\\
State Research
Center\\Institute of Theoretical and Experimental Physics, \\
Moscow, 117218 Russia}

\newcommand{\beq}{\begin{eqnarray}}
 \newcommand{\eeq}{\end{eqnarray}}
\newcommand{\be}{\begin{equation}}
 \newcommand{\ee}{\end{equation}}

 \def\la{\mathrel{\mathpalette\fun <}}
\def\ga{\mathrel{\mathpalette\fun >}}
\def\fun#1#2{\lower3.6pt\vbox{\baselineskip0pt\lineskip.9pt
\ialign{$\mathsurround=0pt#1\hfil ##\hfil$\crcr#2\crcr\sim\crcr}}}

\newcommand{{\SD}}{\rm SD}

\newcommand{{\Mc}}{\mathcal{M}}

\newcommand{\veP}{\mbox{\boldmath${\rm P}$}}
\newcommand{\vep}{\mbox{\boldmath${\rm p}$}}
\newcommand{\veq}{\mbox{\boldmath${\rm q}$}}

\newcommand{\vek}{\mbox{\boldmath${\rm k}$}}

\newcommand{\lan}{\langle}
\newcommand{\ran}{\rangle}

\begin{document}
\maketitle
\begin{abstract} The $\eta$  production in the $(n,n')$
bottomonium transitions $\Upsilon (n) \to \Upsilon (n') \eta, $ is
studied in the method used before  for dipion heavy  quarkonia
transitions. The widths  $\Gamma_\eta(n,n')$ are calculated
without fitting parameters for $n=2,3,4,5, n'=1$.Resulting
$\Gamma_\eta(4,1)$ is found to be large in agreement with recent
data. Multipole expansion method is shown to be inadequate for
large size systems considered.

\end{abstract}

\section{Introduction}
The $\eta$ and $\pi^0$ production in heavy quarkonia transitions
is attracting attention of experimentalists for a long time
\cite{1}. The first result refers to  the $\psi(2S) \to J/\psi
(1S) \eta$ process (to be denoted as $\psi (2,1)\eta$ in what
follows, similarly for $\Upsilon $) with
$\frac{\Gamma_\eta}{\Gamma_{tot}}= (3.09\pm 0.08) \%$ \cite{1}, $
\Gamma_{tot}=337\pm 13$ keV.

For the $\Upsilon (2,1)\eta$ and $\Upsilon (3,1)\eta$ transitions
only upper limits $B<2\cdot 10^{-3}$ and  $B<2.2\cdot 10^{-3}$
were  obtained in \cite{2}  and \cite{3} correspondingly and
preliminary results appeared recently in \cite{4},  $B(\Upsilon
(2,1)\eta) =(2.5\pm 0.7 \pm 0.5)10^{-4}$,  $B(\Upsilon (2,1)
\pi^0) < 2.1\cdot 10^{-4}(90\%$ c.l.) and  $B(\Upsilon (3,1) \eta)
< 2.9\cdot 10^{-4}$ in \cite{4'}.

On theoretical side the dominant approach for both dipion and
single $\eta$ and $\pi$ production is the Multipole Expansion
Method (MEM) (see \cite{5,6} and refs. therein), where it is
asumed that heavy quarks emit two gluons and the latter are
converted into meson(s) by a not clarified mechanism. An essential
requirement for this mechanism in QED is the smallness of the
source size $r_0$ as compared to the wavelength, so that $r_0 k\ll
1$.

In reality both for charmonium and bottomonium transitions
$r_0k\ga 1$,  but it is not this parameter which invalidates MEM
for heavy quarkonia. It appears, that in QCD there is another
important length parameter, the QCD vacuum correlation length
$\lambda$, which makes it impossible to emit freely gluons at
points separated by distance $r, r>\lambda$.

The value of $\lambda$ was found  on the lattice and analytically,
$\lambda\la 0.2 $ fm \cite{8a}. Since r.m.s. radii of  all excited
$c\bar c, b\bar b$ states are larger than 0.5 fm\footnote{In
reality the r.m.s. radius $r_0$ of heavy quarkonia is not small,
e.g. for $\Upsilon(nS)$ it is equal to 0.2 fm, 0.5 fm, 0.7 fm, 0.9
fm, for $n=1,2,3,4,5$,  and for charmonium this radius is even
larger: 0.4 fm and 0.8 fm for $n=1$ and 2 respectively.} all
vacuum gluons there are correlated forming the QCD string and
emission of additional gluons (if any) implies formation of heavy
hybrids. All this is considerd in detail  in Field Correlator
Method (FCM) \cite{8b}.

One  can make an independent check of MEM in application to the
bottomonium level calculation. Here MEM yields nonperturbative
correction to the  levels expressed via gluonic condensate
\cite{6a}. Comparison to the experimental data shows (see
\cite{7a} and Table 1 below) that for all level splittings except
$(2S-1S)$ in bottomonium, MEM prediction is more than 50\% off,
while for charmonium MEM  does not work at all. Thus one concludes
that only at distances below or equal 0.2 fm, MEM can give
reasonable results, while for all states of charmonium and all
excited states of bottomonium (where sizes are much larger than
vacuum correlation length $\lambda$) the application of MEM
is unjustifiable.\\
\vskip .1cm

{\bf Table 1.} Predicted splittings, between spin-averaged levels,
and experiment.

 \vskip .1cm
$$\matrix { {\sl splitting} &MEM [11] & FCM [12]  & {\sl exp.[1]
}
 \cr
& & & \cr {\rm 2S-1S}\,(b\bar {b})&479 & 557& 558\,{\rm MeV} \cr
{\rm 2S-1P}\,(b\bar {b})   &181 &122 & 123\,{\rm MeV} \cr {\rm
3S-2S}\,(b\bar {b})  &4\,570 &348& 332\,{\rm MeV} \cr & & & \cr
{\rm 2S-1S}\,(c\bar {c})&9\,733 &610 & 606\,{\rm MeV} } $$ \vskip
.1cm  \vskip .2cm

A similar failure of MEM is found in applications to dipion
bottomonium  transitions, where using MEM one can fit dipion
spectra in $\Upsilon (2,1) \pi\pi$, but not in $\Upsilon
(3,1)\pi\pi$ and $\Upsilon(4,2) \pi\pi  $ \cite{5,6}. In contrast
to that, FCM as will be discussed below explains both spectra and
$\cos \theta$ dependence for all dipion  transitions in universal
approach with two fixed parameters.

In FCM large distances are under control and not single gluons but
combined effect of all gluons in the string defines the dynamics.

In particular, single eta emission in heavy quarkonia proceeds via
string breaking due to $q\bar q$ pair creation   with simultaneous
emission
 of $\pi$ or $\eta$. The flavor SU(3) violation in $\eta$
production then resides in difference of threshold positions and
wave functions for $ B \bar B$ and $B_s\bar B_s$ ($ D\bar D$ and
$D_s\bar D_s$) intermediate states.

As it is clear, the dynamics of FCM for $\eta$ emission does not
 depend strongly on heavy quark mass, and only sizes of initial and final heavy  quarkonia states
 and intermediate heavy-light mesons enter in the form of overlap
 matrix elements.

 In contrast to that, MEM predicts a strong dependence on the
 heavy quark mass, $\Gamma_\eta (n, n')= O(\frac{1}{m^2_Q})$. In
 addition in \cite{6} a strong suppression of the ratio
 $\Gamma_\eta/\Gamma_{\pi\pi}$ with the growth of the energy release
 $\Delta M = M(n)- M(n'), ~~ \Gamma_\eta/\Gamma_{\pi\pi} \sim
 \frac{p^3_\eta}{(\Delta M)^7}$ is predicted for higher excited
 states of quarkonia and bottomonia,  which does not agree with experiment (see below).

 Using models based on MEM in \cite{5} small ratios of widths \be
\frac{\Gamma(\Upsilon(2,1)\eta)}{\Gamma(\psi(2,1)\eta)} \cong
2.5\cdot 10^{-3} ~~ {\rm
and}~~\frac{\Gamma(\Upsilon(3,1)\eta)}{\Gamma(\psi(2,1)\eta)} =
1.3 \cdot 10^{-3}\label{1}\ee have been predicted, with the model
property   that the bottomonium yields of $\eta$ would be smaller
than those of charmonium; specifically  in the method of \cite{5},
the width is proportional to $p^3/m^2_Q$, so that for $\Upsilon
(4,1) \eta$ the ratio
$\frac{\Gamma(\Upsilon(4,1)\eta)}{\Gamma(\psi (2,1)\eta)}\approx
3.3\cdot 10^{-3}$.

 However recently \cite{8} new BaBar data have
been published on $\Upsilon (4,1)\eta$ with  the branching ratio
\be B(\Upsilon (4,1)\eta) =(1.96\pm 0.06 \pm
0.09)10^{-4}\label{2}\ee and \cite{8} \be
\frac{\Gamma(\Upsilon(4,1)\eta)}{\Gamma(\Upsilon
(4,1)\pi^+\pi^-)}= 2.41\pm 0.40\pm0.12.\label{3}\ee

This latter result  is very large, indeed the corresponding ratio
$\frac{\Gamma(\Upsilon(4,1)\eta)}{\Gamma(\psi (2,1)\eta)}$
 is $\approx 0.4$ and theoretical
estimates  (\ref{1}) from \cite{5} for a similar ratio yields
$3.3\cdot 10^{-3}$.Thus, the experimental  ratio
 is very large as compared to MEM predictions \cite{5,6}.
 All this suggests that
another mechanism can be at work in single $\eta$ production and
below we exploit the approach based on the  Field Correlator
Method (FCM)  recently applied to  $\Upsilon(n,n')\pi\pi$
transitions with $n\leq 3$ in \cite{9,10}, $n\leq 4$ in \cite{11}
and $n=5$ in \cite{12,13}.

 In this paper we confront MEM and FCM and show that recent
 experimental data on single $\eta$ production in $\Upsilon (4S)
 -\Upsilon (1S)$ transition give a strong support to the FCM
 result and cannot be explained in the framework of MEM.

The method essentially expoits  the mechanism of Internal Loop
Radiation (ILR) with light quark loop  inside heavy quarkonium and
has two fundamental parameters -- mass vertices in chiral light
quark pair $q\bar q$ creation $M_{br} \approx f_\pi$ and pair
creation vertex without pseudoscalars, $M_\omega \approx 2\omega$,
where $\omega (\omega_s)$ is the average energy of the light
(strange) quark in the $B(B_s)$ meson. Those are calculated with
relativistic Hamiltonian \cite{7} and considered as  fixed for
all types of transitions $\omega =0.587$ GeV, $\omega_s=0.639$ GeV
(see Appendix 1 of \cite{9} for details).

Any process of heavy quarkonium transition with emission of any
number of Nambu-Goldstone (NG) mesons is considered  in ILR as
proceeding via intermediate states of $B\bar B, B\bar B^*+c.c.,
B_s \bar B_s$ etc. (or equivalently $D \bar D$ etc.) with NG
mesons emitted at vertices.

For one $\eta$ or $\pi^0$ emission one has diagrams shown in
Fig.1, where dashed line is for the NG meson. As shown in
\cite{9,10,11}, based on the chiral Lagrangian derived in
\cite{15}, the meson emission vertex has the structure \be
\mathcal{L}_{CDL} = - i \int d^4 x \bar \psi (x) M_{br} \hat U (x)
\psi (x)\label{4}\ee \be \hat U =\exp \left( i\gamma_5
\frac{\varphi_a\lambda_a}{f_\pi}\right), \varphi_a \lambda_a
=\sqrt{2} \left( \begin{array}{ccc} \frac{\eta}{\sqrt{6}}
+\frac{\pi^0}{\sqrt{2}},& \pi^+,& K^+\\
\pi^-,&\frac{\eta}{\sqrt{6}} -\frac{\pi^0}{\sqrt{2}}, &K^0\\
K^-, &\bar K^0,& -\frac{2\eta}{\sqrt{6}}
\end{array}\right),
\label{5}\ee
 \hspace{-2cm} \small{
\unitlength 1mm 
\linethickness{0.4pt}
\ifx\plotpoint\undefined\newsavebox{\plotpoint}\fi 
\begin{picture}(137,43.25)(0,0)
\put(40.75,31.75){\oval(29,7)[]} \put(110,32.38){\oval(30,6.75)[]}
\put(26,31.5){\circle*{2.12}} \put(54.75,32){\circle*{2.06}}
\put(95.25,33){\circle*{.5}} \put(95,32.5){\circle*{2}}
\put(124.5,32.75){\circle*{1.58}}
\put(40.5,31.63){\oval(33.5,19.25)[]}
\put(109.63,32.38){\oval(33.25,18.75)[]}
\put(25.68,31.68){\line(1,0){.983}}
\put(27.65,31.75){\line(1,0){.983}}
\put(29.61,31.81){\line(1,0){.983}}
\put(31.58,31.88){\line(1,0){.983}}
\put(33.55,31.95){\line(1,0){.983}}
\put(35.51,32.01){\line(1,0){.983}}
\put(37.48,32.08){\line(1,0){.983}}
\put(39.45,32.15){\line(1,0){.983}}
\put(14.25,35){\line(1,0){9.5}} \put(15,27){\line(1,0){9}}
\put(56.75,35.25){\line(1,0){9}}
\put(56.5,27.25){\line(1,0){9.25}}
\multiput(83.25,34.75)(1.21875,.03125){8}{\line(1,0){1.21875}}
\put(83.5,27.25){\line(1,0){9.75}}
\put(126.5,36){\line(1,0){10.5}}
\put(125.25,27.75){\line(1,0){11.25}}
\multiput(124.18,32.93)(.039216,.033497){17}{\line(1,0){.039216}}
\multiput(125.51,34.07)(.039216,.033497){17}{\line(1,0){.039216}}
\multiput(126.85,35.21)(.039216,.033497){17}{\line(1,0){.039216}}
\multiput(128.18,36.35)(.039216,.033497){17}{\line(1,0){.039216}}
\multiput(129.51,37.49)(.039216,.033497){17}{\line(1,0){.039216}}
\multiput(130.85,38.62)(.039216,.033497){17}{\line(1,0){.039216}}
\multiput(132.18,39.76)(.039216,.033497){17}{\line(1,0){.039216}}
\multiput(133.51,40.9)(.039216,.033497){17}{\line(1,0){.039216}}
\multiput(134.85,42.04)(.039216,.033497){17}{\line(1,0){.039216}}
\put(37.75,15.25){\makebox(0,0)[cc]{( a )}}
\put(104,17.25){\makebox(0,0)[cc]{( b )}}
\end{picture}}
\vspace{-0.5cm}

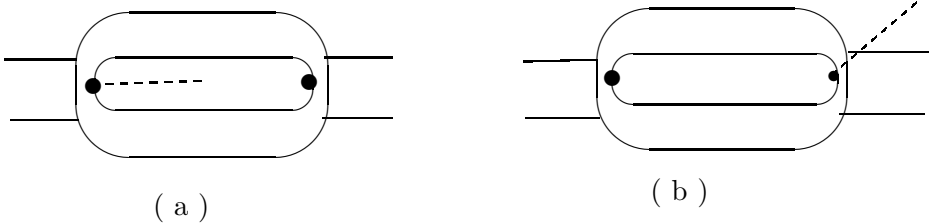
 Fig.1 Single eta
production (dashed line) from
 $\Upsilon(n)BB^*$ vertex (a), and
$BB^*\Upsilon(n')$ vertex (b).

\vspace{1cm}

The lines (1,2,3) in the $\hat U$ matrix  (\ref{2}) refer to
$u,d,s$ quarks and hence to the channels $B^+B^-, B^0\bar B^0,
B^0_s \bar B^0_s$ (and to the corresponding channels with $B^*$
instead of $B$). Therefore the emission of a single $\eta$  in
heavy quarkonia transitions requires the flavour $SU(3)$ violation
and resides  in our approach in the difference of channel
contribution $B\bar B^*$ and $B_s\bar B_s^*$, while the $\pi^0$
emission is due the difference of $B^0\bar B^{0*}$ and $B^+
B^{-*}$  channels (with $B\to D$ for charmonia).

The paper is devoted to the  explicit calculation of single $\eta$
 emission widths in bottomonium $\Upsilon (n,1)
\eta$ transitions with $n=2,3,4,5.$ Since theory has no fitting
parameters (the only ones, $M_\omega$ and $M_{br}$ are fixed by
dipion transitions) our predictions depend only on the overlap
matrix elements, containing wave functions of $\Upsilon (nS)$, $B,
B_s, B^*, B^*_s$. The latter have been computed previously in
relativistic Hamiltonian technic in \cite{7} and used extensively
in dipion transitions in \cite{11,12,13}.

The paper is organized as follows. In section 2 general
expressions for process amplitudes are given;  in section 3
results of calculations are presented and discussed and a short
summary and prospectives are given.

\section {General formalism}

The process of single NG boson emission in bottomonium transition
is described  by two diagrams depicted in Fig.1,  (a) and (b)
which can be written according to the  general formalism of FCM
\cite{9,11,12} as (we consider $\eta$ emission), see Appendix for
more detail,

\be \Mc=\Mc_\eta^{(1)}+\Mc_\eta^{(2)}; \Mc_\eta^{(i)}=
\Mc^{(i)}_{B_sB^*_s} - \Mc^{(i)}_{BB^*} , i=1,2\label{6}\ee For
the diagram of Fig. 1(a) the amplitude for intermediate $BB^*$ or
$B_s B^*_s$ state can be written as
 \be
\Mc_\eta^{(1)}=\int\frac{J^{(1)}_n (\vep, \vek) J_{n'}
(\vep)}{E-E(\vep)} \frac{d^3\vep}{(2\pi)^3},\label{7}\ee while
 $\Mc_\eta^{(2)}$, corresponding to the diagram of Fig.1(b),
  has the same form, but without  NG boson energy in the
 denominator of (\ref{7}). The overlap integrals of $\Upsilon (nS)$ and $BB^*$
 wave functions with emission of $\eta$ with momentum $\vek$ are
 denoted by $J_n^{(i)} (\vep,\vek)$, the corresponding integrals
 without $\eta$ emission are given by $J_{n'}(\vep)$.

Finally we define  all quantities in the denominator of (\ref{7});
in $\Mc^{(1)}_{BB^*}$  the denominator is \be
E-E(\vep)=M(\Upsilon(nS))-(\omega_\eta+M_B+M^*_B+
\frac{\vep^2}{2M_B} +\frac{(\vep-\vek)^2}{2M^*_B})\equiv-\Delta
M^*- \omega_\eta-E(\vep, \vek).\label{11}\ee
\newcommand{{\Lc}}{\mathcal{L}}
For $\Mc_\eta^{(2)}$ one omits $\omega_\eta$ and $\vek$ in
(\ref{11}). Finally  after taking Dirac traces in amplitudes and
accounting for the $p$-wave of emitted $\eta$ one can represent
the matrix element $\Mc^{(i)}_\eta$ as follows: \be \Mc_\eta^{(1)}
= e_{ii'l} k_l \left( \frac{1}{\omega^3_s}\Lc^{(1)}_s
-\frac{1}{\omega^3} \Lc^{(1)} \right)\label{12}\ee
 Indices $i'i$ in
$e_{i'il}$ in (\ref{12}) refer to the $\Upsilon(n'S)$ and
$\Upsilon (nS)$ polarizations respectively.

\be \Mc_\eta^{(2)} = e_{ii'l} k_l \left(
\frac{1}{\omega^3_s}\Lc^{(2)}_s -\frac{1}{\omega^3} \Lc^{(2)}
\right) \label{13}\ee with  $\Lc^{(i)}, \Lc_s^{(i)}$ being
integrals of overlap factors $J_n(\vep,\vek)J_{n'}(\vep).$

The width of the $\Upsilon (n,n') \eta$ decay is obtained  from
$|\Mc|^2$ averaging over vector polarizations as \be \Gamma_\eta
=\frac13 \sum_{i,i'} |\Mc|^2 d\Phi = \frac{2 k^2}{3}  d\Phi \left|
\left(\frac{\Lc_s^{(1)}}{\omega^3_s}-\frac{\Lc^{(1)}}{\omega^3}\right)+
\left(\frac{\Lc_s^{(2)}}{\omega^3_s}-\frac{\Lc^{(2)}}{\omega^3}\right)\right|^2\label{16}\ee
where the phase space factor
$d\Phi=\frac{d^3k}{(2\pi)^3}2\pi\delta(M(\Upsilon(n))-
M(\Upsilon(n'))-\omega_\eta-\frac{k^2}{2M(\Upsilon(n'))})$.

Introducing the average $\bar \omega=\frac12 (\omega_s+\omega)$,
one can rewrite (\ref{16}) as $$ \Gamma_\eta=
\left(\frac{M_{br}}{f_\eta}\right)^2\left(\frac{M_{\omega}}{2\bar
\omega}\right)^2\zeta_\eta \frac{k^3}{\bar \omega^4}
e^{-\frac{k^2}{2\beta^2_2}}\left| \left(\frac{\bar
\omega}{\omega_s}\right)^3 \Lc_s^{(1)}- \left(\frac{\bar
\omega}{\omega}\right)^3\Lc^{(1)}  +\right.$$ \be\left.
\left(\frac{\bar \omega}{\omega_s}\right)^3 \Lc_s^{(2)}-
\left(\frac{\bar \omega}{\omega}\right)^3\Lc^{(2)}
\right|^2\label{17}\ee with $\zeta_\eta=\frac{16}{9\pi N^2_c}\cong
0.063$.

One can see from the general structure of $\Gamma_\eta$, that the
main effect comes from the difference $\left| \left(\frac{\bar
\omega}{\omega_s}\right)^3-\left(\frac{\bar
\omega}{\omega}\right)^3\right|\approx |0.882 -1.139|\approx
0.257$, and  from  the difference of
$|\Lc_s^{(i)}-\Lc^{(i)}|\leq0.05 $.

\section{Results and discussion}

We consider here the single $\eta$ emission in bottomonium
transitions $\Upsilon (n,1) \eta$ with $n=2,3,4,5$. The
corresponding values of $\Delta M^*, \Delta M^*_s, \omega_\eta, k$
are given in the Table 2.

{\bf Table 2.}  Mass parameters of $\Upsilon (n,n')\eta$
transitions (all in GeV, $k$ in GeV/c).

 \begin{center}
\vspace{3mm}

\begin{tabular}{|l|l|l|l|l|} \hline
&&&&\\

$(n,n')$& 2,1&3,1&4,1&5,1\\
\hline  $\Delta M^*$ & 0.582&0.25&0.026&-0.26\\
\hline  $\Delta M^*_s$ & 0.757&0.425&0.20&-0.08\\
\hline $\omega_\eta$ & 0.562& 0.87&1.075&1.325\\
\hline  $k$& 0.115&0.674&0.923&1.20\\

\hline
\end{tabular}

\end{center}

The resulting values of $\Gamma_\eta (n, n')$ have been computed
as in (\ref{17}) with $\omega=0.587$ GeV and $\omega_s=0.639$ GeV,
calculated earlier in \cite{7}, see Table 4 of \cite{9}, and with
wave functions  fitted to the  realistic wave functions in
\cite{11}, (set I), while in set II parameters of the
$\Upsilon(nS)$ wave function were changed by 10-15\%.

Results of calculations are given in Table 3, where we have put $M_{br} \approx f_n$ and
 $\left( \frac{M_\omega}{2\omega}\right)^2 =\frac12$ to explain the decay $\Gamma_{BB} (4S) =\Gamma_{exp}$ ~(see
 \cite{11} for details). \\

\vspace{3mm}

 {\bf Table 3.}  Values of $\Gamma_\eta(n,n') $  (in
keV) calculated using Eq. (\ref{17}) with  $\left(
 \frac{M_{br}}{f_n} \frac{M_\omega}{2\omega}\right)^2 =\frac12$  $ vs$ experimental data
$\Gamma^{\exp}_\eta (n,n')$  (in keV).

 \begin{center}
\vspace{3mm}

\begin{tabular}{|c|c|c|c|c|} \hline
&&&&\\

$(n,n')$& (2,1)& (3,1)&(4,1)&(5,1)\\
\hline $\Gamma_\eta$, set I & 2.50$\cdot 10^{-2}$& 1.45&
0.9&3.5\\\hline$\Gamma_\eta$, set II &  $ 1.4\cdot 10^{-2}$&
3.6$\cdot 10^{-3}$ &1.9 &1.3\\\hline $\Gamma^{exp}_\eta$
&(0.8$\pm0.3)\cdot 10^{-2}$&$<5.8\cdot 10^{-3}$&$4.0\pm 0.6$&-\\
&[4]&[5]&[13]&\\
\hline
\end{tabular}

\end{center}

Looking at the Table 3, one can see, that there is an order of
magnitude agreement with experiment for the set II. Indeed, the
factor ${\left(\frac{M_{br}}{f_\pi}\right)^2
\left(\frac{M_\omega}{2\bar \omega}\right)^2}$  can be estimated
from $\Upsilon (n, n')\pi\pi$ transitions studied in
\cite{9}-\cite{13} to be roughly in the range $[\frac12, 2]$.  On
theoretical side our formulas (\ref{7})-(\ref{16}) automatically
produce the width $\Gamma_\eta(n,n')$ of the order of $O(1$ keV),
for all $(n,1)$ transitions except for (2,1),  where a small phase
space factor $k^3$ gives two orders of magnitude suppression of
$\Gamma_\eta(2,1)$, and for $\Gamma_\eta(3,1)$, which is highly
sensitive to the form of wave  function  (cf. set I and  set II).
For the $\Gamma_\eta(5,1)$ one obtains a 7 keV value, which is
however small as compared with the $\Gamma_{\pi\pi}(5,1)$, the
latter being $O(1$ MeV). For $\Gamma_\eta (4,1)$ and $
{\Gamma_{\pi\pi}(4,1)}$ from \cite{11} the calculated ratio is
$R_{\eta/\pi\pi} \equiv \frac{\Gamma_\eta (4,1)}{\Gamma_{\pi\pi}
(4,1)} \cong 3\left(\frac{M_\omega}{2\bar \omega}\right)^2
\left(\frac{f_\pi}{M_{br}}\right)^2\approx 1.5$, which roughly
agrees with experimental value $R^{\exp}_{\eta/\pi\pi}= 2.41\pm
0.40 \pm 0.12$ \cite{8}.

To compare our  results with MEM predictions and experiment, we
list in Table 4 the ratios
$\frac{\Gamma(\Upsilon{(n,1)}\eta)}{\Gamma(\psi(2,1)\eta)}\equiv
X$ for $n=2,3,4,5$ (the MEM numbers for $n=4,5$ are obtained from
$n=3$ using scaling $X=O(p^3)$ \cite{5}.\\

\vspace{3mm}

 {\bf Table 4.}  Values of the ratio  $\frac{\Gamma(\Upsilon{(n,1)}\eta)}{\Gamma(\psi(2,1)\eta)}\equiv
X$ for $n=2,3,4,5$ from MEM \cite{5}, experiment and the present
paper (for the latter  we take $\Gamma_{\exp} (\psi (2,1)\eta)$ in
the denominator  and two sets from Table 3):

 \begin{center}
\vspace{3mm}

\begin{tabular}{|c|c|c|c|c|} \hline
&&&&\\

$n$& 2& 3&4&5\\
\hline MEM& $0.25\cdot 10^{-2}$ &$ 0.13\cdot 10^{-2}$& 0.33$\cdot
10^{-2}$ &$ 0.73\cdot 10^{-2}$\\
\hline exper.
&(0.8$\pm0.2)\cdot 10^{-3}$&$<0.58\cdot 10^{-3}$&$0.40\pm 0.06$&-\\
&[4]&[5]&[13]&\\
\hline present paper& $(1.4\div 2.5)\cdot 10^{-3}$ & $(0.36\cdot
10^{-3}\div 0.15)$& $0.9\div 1.9$& $0.13\div 0.35$\\\hline

\end{tabular}

\end{center}

 To check stability of our results,
we have used for the wave function of $B_s$  the realistic wave
function different from that of $B$. As a result  one obtains for
$\frac{\Gamma_\eta(n,1)}{\left(\frac{M_{br}}{f_\pi}\right)^2\left(\frac{M_{\omega}}{2\omega}\right)^2}$
the values $(2.74\cdot 10^{-2}$; 1.13; 0.44; 7.3) keV  for
$n=2,3,4,5$ respectively. To understand why $\Gamma_\eta (.1)$ is
so sensitive to the form of the wave function (cf. results for the
sets I and II), we have varied the shape  of wave functions of
$\Upsilon (nS)$ with $n=2,3,4,5$ (explicitly the parameter
$\beta_1$) in the range $ \xi=\frac{\beta_1(n)}{\beta_1^{opt}
(n)}=0.8\div 1.2$ of the optimal value, reproducing the realistic
wave function. We have obtained the   dependence of $\Gamma_\eta
(n,1)$ on $\xi$ for $n=2,3,4,5$. As a result  $\Gamma_\eta (2,1) $
is rather stable, whereas $\Gamma_\eta (3,1)$ has minimum (almost
zero) for $\xi= 1.17$. This fact explains results of sets I and
II; while set I produces normal (and nearly maximal) values for
$\Gamma_\eta (n,1)$ of the order of 1 keV for $n>2$, in case of
set II the parameter $\beta_1$ of $\Upsilon (3S)$ roughly
corresponds to the minimum of $\Gamma_\eta$. In this way  doing
comparison with experiment one obtains  possibly an instrument for
a precision study of wave functions of excited bottomonium states.

Summarizing, we have calculated  the single $\eta$ production
width $\Gamma_\eta(n,n')$ for  $\Upsilon (n,1) \eta$ transitions
with $ n=2,3,4,5$. We have found that $\Gamma_\eta(4,1)$ are of
the order of and larger than  $\Gamma_{\pi\pi} (4,1)$. This fact
is in agreement with the latest measurements in \cite{8} of
$\Gamma_\eta^{\exp}(4,1)$. We have shown that $\Gamma_\eta (n,1)$,
$n=4,5$  is large $(\sim O(1$ kev)) for  typical (realistic)
parameters of $\Upsilon (nS)$ wave functions, but can occasionally
drop near zero for slightly varied form of wave function, as it
happens for $n=3$. This high sensitivity  is connected to the
oscillating  character of excited bottomonium wave
functions\footnote{A similar phenomenon of cancellation in matrix
elements was mentioned in \cite{5}}. Our calculations do not
contain fitting parameters; the only two parameters $M_{br},
M_\omega$ are fixed by previous comparison with dipion data.  One
should stress that $\eta$ production in bottomonium is not
suppressed in our approach as compared to $\eta$ production in
charmonium transitions. This is in contrast with the results of
method of \cite{5,6}. We have given arguments why the dipion
transitions in  high excited states of heavy quarkonia as well as
single $\eta$ and $\pi$ emission cannot be reliably calculated
within the MEM method of \cite{5,6}, widely used now. As it is
seen in Tables 3 and 4 the sequence of experimental data
\cite{4},\cite{4'}, \cite{8} contradict predictions of MEM.
Recently a new calculation was done of $\Gamma_\eta (n, n')$ in
\cite{21} where also $B\bar B^*$ etc. intermediate states were
taken into account as well as in our approach. The authors however
did not use wave functions of hadrons involved, but  rather
exploited fitted coupling constants and formfactors, and specific
form of matrix elements, which makes it difficult to compare with
our method directly.

The authors are grateful to S.I.Eidelman for constant support and
helpful comments.

 The financial support of  grants RFFI
06-02-17012,  06-02-17120 and NSh-4961.2008.2 is gratefully
acknowledged.

{\bf Appendix 1}\\

{\bf Matrix element of single $\eta$ emission}\\

 \setcounter{equation}{0} \def\theequation{A.\arabic{equation}}

According to the general theory in \cite{9,11},  the matrix
elements $\Mc^{(1)}_{\eta}, |Mc_\eta^{(2)}$ for
$\Upsilon(n,n')\eta$ corresponding to diagrams of Fig.1, (a)  and
(b) respectively, can be written as \be \Mc_\eta^{(1)} (n)
=\frac{M_{br} M_\omega}{f_\eta N_c\sqrt{2\omega_\eta}} \int
\frac{d^3p}{(2\pi)^3} \sum_{n_2,n_3} \frac{J_{n
n_2n_3}{(\vep,\vek)} J^+_{n'n_2n_3}
(\vep)}{E-E_{n_2n_3}(\vep)}.\label{A1}\ee Here $n_2, n_3$ are
channels of intermediate state, with e.g. $n_2 =B, B^*, B_s,
B_s^*, ...,$ we omit indices $n_2, n_3$ and write \be Jn (\vep,
\vek) =\int \bar y^{(\eta)}_{n23} \frac{d^3\veq}{(2\pi)^3} \tilde
\Psi_n (c\vep -\frac{\vek}{2} +\veq) \tilde \psi_{n_2} (\veq)
\tilde \psi_{n_3} (\veq-\vek)\label{A2}\ee where $\tilde \Psi_n,
\tilde \psi_{n_i}$ are momentum space  wave functions of $\Upsilon
(nS)$ and $B (B^*)$ mesons respectively.

The vertex factor $\bar y^{(\eta)}_{123}$ is calculated in the
same way as in \cite{9}, namely from the Dirac trace of the
projection operators for the decay process in our case this is
$\Upsilon (nS)  \to BB^*\eta$. Identifying the creation operators
as $\bar \psi_b \gamma_i \psi_b, \bar  \psi_b\gamma_5 \psi_n, \bar
\psi_b \gamma_k \psi_n ,  ~ n=u,d,s a$ and extracting vertex of
$\eta$ creation from the Lagrangian $ \Delta\Lc =- \int \bar
\psi_n \hat U  M_{br} \psi_n d^4 x$ which gives $i\frac{ M_{br}
\bar \psi_n \gamma_5 \hat \lambda \psi_n}{f_n
\sqrt{2\omega_\eta}}$, with $\hat \lambda=\frac{1}{\sqrt{3}}
\left(\begin{array}{lll} 1&&\\&1&\\&&-2 \end{array}\right)$, one
has for the decay process (cf. Appendix 1 of \cite{9}) \be
G(\Upsilon \to BB^*\eta) =tr [\gamma_i S_b (u,w) \gamma_5 S_{\bar
n} (w, x) \gamma_5S_n (x, w) \gamma_k S_{\bar b}
(w',v)]\label{A3}\ee

As  shown in \cite{9}, appendix 1 and 2, the quark Green's
functions can be split into two factors $S(x,y) =\Lambda^\pm
G(x,y)$, with the projection operators  $\Lambda^\pm_k =\frac{
m_k\pm \omega_k \gamma_4 \mp i p_i ^{(k)} \gamma_i}{2\omega_k} ,
k=b,n$ and  the scalar part $G(x,y)$, where spins are present only
in spin-depdndent interaction and treated as corrections. Here
$\omega_k$ is the average energy of quark in given meson. Hence
one is brought to the spin factor $Z$. \be Z=tr (\gamma_i
\Lambda^+_b \gamma_5 \Lambda^-_n \gamma_5 \Lambda_n^+ \gamma_k
\Lambda^-_b)\label{A4}\ee
 which is equal to
 \be
 Z=\frac{m^2_b+\Omega^2}{4\omega^2\omega^2} ((\omega^2
 -\vep^q\vep^{\bar q}) \delta_{ik} - p^q_i p^{\bar q}_k + p^q_k
 p^{\bar q}_i).\label{A5}\ee

 Here  $\Omega, \omega$ are average energies of $b$ and $n$ quark
 in $B$ or $B^*$. One can identify the momenta of $B$ and $B^*$ as
 $\veP_1 =\vep$ and $\veP_2 = -\vep-\vek$, then $\veq$ in
 (\ref{A2}) can be expressed as
 \be \vep_{\bar q} =- \veq +\frac{\omega}{\omega +\Omega} \vep,
 ~~\vep_q = \veq -\frac{\omega}{\omega+\Omega} \vep - \vek
 \frac{\Omega+2\omega}{\Omega+\omega},\label{A6}\ee
 and $Z$ is (we put $m_b \cong\Omega)$
 \be
 Z=\frac{1}{2\omega^2}(-\vek\veq\delta_{ik} +k_iq_k+
 k_kq_i).\label{A7}\ee
 It is important, that we are looking for the $P$-wave of emitted
 $\eta$, and hence for $P$ wave of relative $BB^*$ motion, hence
 the integral (\ref{A2}) should yield the term $\vep\vek$. This
 indeed happens, when one  approximates $\tilde \Psi_n, \tilde
 \psi_n$ as series of oscillator wave functions and (\ref{A2}) has
 the form
 \be
 I_n (\vep, \vek) =\bar y_{n23}^\eta
 e^{-\frac{\vep^2}{\Delta_n}-\frac{\vek^2}{4\beta^2_2}}~^{(0)}I_n
 (\vep).\label{A8}\ee
In the process of $d\veq$ integration in (\ref{A2}) one changes
the integration  variable $q_i \to q'_i - u_n +O(k_i)$ with
$=\beta_2, \Delta_n$ are  oscillator parameters, found by $chi^2$
procedure.

Thus result of $d^3q $ integration yields \be\bar y^\eta_{123} =
\frac{u_n}{2\omega^2} (-\vek\vep \delta_{ik} -k_i p_k+ K_k
p_i).\label{A9}\ee

In an analogous way one obtains for $J_{n'}(\vep)$ in (\ref{A1})
the form \be J_{n'} (\vep) = \bar y_{n'23}^{(\eta)}
e^{-\frac{\vep^2}{\Delta_{n'}}}~^{(1)} I_{n'} (\vep)\label{A10}\ee
and $\bar y^{(\eta)}_{n'23}$ is obtained from the Dirac trace for
the process $B\bar B^*\to \Upsilon (n'S')$, \be Z_2 (BB^*)
=\frac{1}{2\omega} e_{i'kl} (-2q_l +\frac{2\omega}{\omega+\Omega}
p_l)\label{A11}\ee and the result of integration over $d^3q$
yields in (\ref{A10})\be  \bar y_{n'23}^{(\eta)} =- e_{i'kl}
\frac{p_l}{\omega}.\label{A12}\ee

Here $i'$ is the polarization of  $\Upsilon(n'S)$ (represented by
$ \bar\psi_b \gamma_{i'} \psi_b$) and $k$ as in (\ref{A9}) is the
polarization of $B^*$. Averaging over angles of  $\vep$ one
obtains\be \lan \bar y^{(\eta)}_{n23} \bar y^{(\eta)}_{n'23}\ran_p
= \frac{u_n}{3} \frac{\vep^2}{\omega^3}
(e_{i'il}k_l)\label{A13}\ee and finally one writes as in
(\ref{12}) \be \mathcal{M}_n^{(1)} (n,n') =\frac{M_\omega M_{br}
u_n 2e_{i'il} k_l}{f_n\sqrt{2\omega_\eta} \sqrt{3}\cdot 3}
\left(\frac{\mathcal{L}^{(1)}}{\omega^3}
-\frac{\mathcal{L}^{(1)}_s}{\omega^3_s}   \right)
e^{-\frac{k^2}{4\beta^2_2}}.\label{A14}\ee

For $ \mathcal{M}_n^{(2)} (n,n')$ one can use time inversion and
interchange indices $i,i'$ and change sign of $\vek$, obtaining in
this way Eqs. (\ref{12}) and (\ref{13}) of the main text. For the
intermediate state of $B^*\bar B^*$ the summation over
polarizations of $B^*$ yields a net zero result, therefore we are
left with only $(B\bar B^*+ B^*\bar B)$ intermediate state.

\end{document}